\documentclass[twocolumn,showpacs,preprintnumbers,amsmath,amssymb,prb]{revtex4}
\usepackage[dvips]{color,graphics,epsfig,rotating}
\usepackage{graphicx}
\usepackage{dcolumn}
\usepackage{bm}
\usepackage{color}

\newcommand\eql[2] 
{
\begin{equation}\label{#1}
\begin{split}
#2
\end{split}
\end{equation}
}

\DeclareMathOperator{\erfc}{erfc}
\newcommand\Ecut {E_{\textrm{cut}}}
\newcommand\Econv {E_{\textrm{conv}}}
\newcommand\ME[3]      {\langle{{#1}}|{{#2}}|{{#3}}\rangle} 
\newcommand\ket[1]     {|{{#1}}\rangle}
\newcommand\bra[1]     {\langle{{#1}}|}
\newcommand\braket[2]  {\langle{{#1}}|{{#2}}\rangle}
\newcommand\bohr       {\,a_0}
\newcommand\PsiGS      {\Psi_0}
\newcommand\Dc[1]      {c_{{#1}}^{}}
\newcommand\Cc[1]      {c_{{#1}}^\dagger}
\newcommand\Half       {\frac{1}{2}}
\newcommand\Gvec       {\mathbf{G}}
\newcommand\Qvec       {\mathbf{Q}}
\newcommand\Rvec       {\mathbf{R}}
\newcommand\rvec       {\mathbf{r}}
\newcommand\Aop        {{\hat{A}}}
\newcommand\Bop        {{\hat{B}}}
\newcommand\Hop        {{\hat{H}}}
\newcommand\Kop        {{\hat{K}}}

\newcommand\Veiop      {\hat{V}_{\textrm{ei}}}
\newcommand\Veeop      {\hat{V}_{\textrm{ee}}}
\newcommand\VeiLop     {\hat{V}_{\textrm{ei,L}}}
\newcommand\VeiNLop    {\hat{V}_{\textrm{ei,NL}}}
\newcommand\VL         {V_{\textrm{L}}}
\newcommand\VNL        {V_{\textrm{NL}}}
\newcommand\Vii        {V_{\textrm{II}}}
\newcommand\HSop       {{\hat{b}}}

\newcommand\GAUSSIAN[1][]{{\footnotesize{GAUSSIAN#1}}}
\newcommand\ABINIT     {{\footnotesize{ABINIT}}}
\newcommand\ECP        {OAL} 
\newcommand\ECPpsp[1]  {\ECP{} pseudopotential#1} 
\newcommand\IP         {\mathrm{IP}}
\newcommand\IIP        {\mathrm{IIP}}
\newcommand\De         {D_e}

\definecolor{Green}{rgb}{0.2,0.96,0.2}
\definecolor{Remarks}{rgb}{1,0.3,0.3}
\definecolor{Extra}{rgb}{0.2,0.2,1}
\definecolor{Blue}{rgb}{0.2,0.3,1}
\definecolor{Black}{rgb}{0,0,0}

\newcommand\COMMENTED[1] {}

\newcommand\FIGDIR[1]    {}

\begin{document}

\title{Phaseless auxiliary-field quantum Monte Carlo calculations
with planewaves and pseudopotentials---applications to atoms and molecules}

\author{Malliga Suewattana}
\altaffiliation[Present address: ]{Oak Ridge National Laboratory,
Oak Ridge, TN 37831-6030}
\author{Wirawan Purwanto}
\author{Shiwei Zhang}
\author{Henry Krakauer}
\author{Eric J. Walter}
\affiliation{Department of Physics, College of William and Mary, Williamsburg,
Virginia 23187-8795}

\date{\today}

\begin{abstract}

The phaseless auxiliary-field quantum Monte Carlo (AF QMC) method [S. Zhang and H. Krakauer, {\it
Phys. Rev. Lett.} {\bf 90}, 136401 (2003)] is used to carry out a
systematic study of the dissociation and ionization energies of
second-row group 3A-7A atoms and dimers, Al, Si, P, S, Cl.  In
addition, the P$_2$ dimer is compared to the third-row As$_2$ dimer,
which is also triply-bonded.
This method projects the many-body
ground state by means of importance-sampled random walks in the space of Slater determinants.
The Monte Carlo phase problem, due to the electron-electron Coulomb interaction, is 
controlled via the phaseless approximation, with a trial wave function  $|\Psi_T\rangle$.
As in previous calculations, a mean-field single Slater determinant is used
as $|\Psi_T\rangle$. 
The method is formulated in the Hilbert space defined by any chosen one-particle basis.
The present 
calculations use a planewave basis under periodic boundary conditions with
norm-conserving pseudopotentials.
Computational details of the planewave AF QMC method are presented.
The isolated systems chosen here allow a systematic study of the various algorithmic issues.
We show the accuracy of the planewave method and discuss its convergence with
respect to parameters such as the supercell size and planewave cutoff.
The use of standard norm-conserving pseudopotentials in the
many-body AF QMC framework is examined.

\end{abstract}

\COMMENTED{
\begin{verbatim}
PLANEWAVE QMC PAPER 1: ATOMS & MOLECULES (draft 8) HK->SZ->WP MODIFICATIONS
CVS $Id: QMC.tex,v 1.17 2007/02/05 04:51:21 wirawan Exp $
---------------------------------------------------------------------------


\end{verbatim}
}


\maketitle

\section{Introduction}

Achieving accurate solutions of the electronic many-body Schr\"odinger equation 
is a challenging problem for calculations of the properties of real materials.
For many systems, density functional theory (DFT), in a variety of approximations,
has been applied with great success. In DFT, the many-body
interactions are replaced by a single particle interacting with the
mean-field generated by the other particles, similar in spirit to the
Hartree-Fock (HF) method. Unfortunately, these methods have well-known
limitations and often fail at describing the properties of materials with
large electron-electron correlation.

A more accurate approach is the quantum Monte Carlo (QMC) method
\cite{Ceperley1980, Reynolds1982, Foulkes2001,sz-hk}, which
has been shown to be among the most effective methods for
many-electron problems. 
Unlike other
correlated methods, QMC calculation times scale as a low power of the system size
 \cite{WilliamsonEtAl}. 
The fixed-node diffusion Monte Carlo (DMC) approach, which samples the many-body 
wave function in real space, has been the most widely used 
QMC method in electronic structure calculations \cite{Reynolds1982,Foulkes2001}.

The recently developed phaseless auxiliary-field quantum Monte Carlo (AF QMC)
method \cite{sz-hk} is an
alternative and complementary QMC approach, which samples the many-body wave function
in the space of Slater determinants. 
This method has several attractive features. The fermionic antisymmetry
of the wave function is automatically accounted for, since it is sampled
by Slater determinants. This provides a different route to controlling the sign problem
\cite{Ceperley_sign,kalos91,Zhang1999_Nato} from fixed-node DMC, which has shown promise in reducing the
dependence of the systematic errors on the trial wave function
\cite{Al-Saidi_TMO, Al-Saidi, Al-Saidi-06JCP2}.
The orbitals in the Slater determinants are expressed 
in terms of a chosen single-particle basis (\textit{e.g.}, planewaves, Gaussians, etc.),
so AF QMC shares much of the same computational machinery with DFT and other 
independent-particle type methods. AF QMC can thus straightforwardly 
incorporate many of the methodological advances from mean-field methods 
(such as pseudopotential and fast Fourier transforms) while
systematically improving on mean-field accuracy.

Using a planewave basis, tests of the phaseless 
AF QMC method for a few simple atoms and molecules \cite{sz-hk, AFQMC-CPC2005} as well as for
the more correlated TiO and MnO molecules \cite{Al-Saidi_TMO} 
yielded excellent results.
More systematic applications of the phaseless AF QMC method to atoms and
molecules have been
carried out using Gaussian basis sets. These include all-electron calculations 
for first-row systems \cite{Al-Saidi} 
and effective-core
potential calculations in post-$d$ group elements \cite{Al-Saidi-06JCP2}. 
The results also showed excellent agreement with near-exact quantum chemistry
results and/or experiment. 

The planewave AF QMC method is well-adaped for correlated calculations of 
extended bulk systems, where planewave based methods have been the standard choice in 
traditional electronic structure calculations.
It is therefore important to systematically study its algorithmic issues and 
to characterize its performance.
In this paper, we use the planewave phaseless AF QMC method to carry
out a systematic study of the dissociation and ionization energies of
second-row atoms and dimers in Group 3A-7A, namely Al, Si, P, S, Cl.  The
interesting case of the triply-bonded P$_2$ dimer is also compared to
the third-row As$_2$ dimer. 
The principal goal of this study is to further benchmark the
AF QMC method across more systems and across different basis sets and
to compare the 
results with those from other methods and experiment. 
While the use of localized basis sets, such as
Gaussians, is generally more efficient for isolated atoms and molecules,
it is straighforward to apply planewave methods using periodic boundary conditions
and large supercells.
Planewave methods have several desirable features.
A planewave basis provides an
unbiased representation of the wave functions, since convergence to
the infinite basis limit is controlled by a single parameter, the
kinetic-energy cutoff $\Ecut$. Planewaves are algorithmically
simple to implement, and operations with planewaves can be made very
efficient, using fast Fourier techniques as in DFT methods.  
To
keep the planewave basis size tractable, pseudopotentials must be used to 
eliminate the highly localized core electron states and to produce relatively
smooth valence wave functions.
The present choice of isolated atomic and molecular systems permits
direct comparisons with Gaussian-based AF QMC and with well-establised
quantum chemistry all-electron and pseudopotential calculations.

The remainder of this paper is organized as follows. In Section \ref{sec:compt},
we describe computational details of the phaseless AF QMC method in the planewave-pseudopotential framework. 
In Section \ref{sec:convergence},
we discuss 
calculation parameters such as supercell (simulation
cell) size, cutoff energy, and AF QMC time step size, and the algorithm's
convergence behavior with respect to these parameters.
Section \ref{sec:results} presents the calculated
dissociation and ionization energies and comparisons with other theoretical results
and with experiment. In Section \ref{sec:discussion}, we discuss systematic 
errors due to the use of the phaseless approximation and use of norm-conserving pseudopotentials. 
We then conclude with some general remarks in Section \ref{sec:summary}.

\section{Planewave AF QMC method: Computational Details}
\label{sec:compt}

\subsection{Hamiltonian}
\label{sec:Hamiltonian}

It is convenient to express, within the Born
Oppenheimer approximation, the electronic Hamiltonian in second
quantized form in terms of a chosen orthonormal one-particle basis
\begin{equation}
\label{eq:H}
    \Hop
  = \sum_{ij}^M {H^{(1)}_{ij} \Cc{i}\Dc{j}}
  + \Half \sum_{ijkl}^M
    {H^{(2)}_{ijkl} \Cc{i}\Cc{j}\Dc{l}\Dc{k}}
  + \Vii
    \,,
\end{equation}
where $M$ is the number of basis functions, $c_i^\dagger$ and $c_i$
are the corresponding creation and annihilation operators, and
the electron spins have been subsumed in the summations. 
$H^{(1)}_{ij}$ and $H^{(2)}_{ijkl}$ are the one- and two-body matrix
elements, and $\Vii$ is the classical Coulomb interaction
of the point ions \cite{Yin_1982}.
Atomic units are used throughout this paper.
We use periodic boundary conditions and a planewave basis
\begin{equation}
   \braket{\rvec}{\Gvec}
   \equiv
   \ME{{\rvec}}{\,\Cc{\Gvec}}{0}
 = \frac{1}{{\sqrt \Omega}}\exp (i{\Gvec} \cdot {\rvec})
   \, ,
\end{equation}
where $\Omega$ is the volume of the simulation cell and ${\Gvec}$ is a
reciprocal lattice vector.  As in planewave-based density functional
calculations, the number of planewaves $M$ in the basis is determined
$G^2/2 \le \Ecut$, where $\Ecut$ is the cutoff kinetic energy.

The one-body operators in the Hamiltonian include the kinetic energy,
\begin{equation}
\label{KE}
    \Kop
  = \Half \sum_{\Gvec} {G^2}\, \Cc{\Gvec} \Dc{\Gvec}
    \,,
\end{equation}
and nonlocal pseudopotential, which describes the electron-ion interaction
\eql{V_ei}
{
    \Veiop
& = \sum_{\Gvec,\Gvec'} \VL({\Gvec} - {\Gvec'})\, \Cc{\Gvec}\Dc{\Gvec'}
\\
&\,
  + \sum_{\Gvec,\Gvec'} \VNL({\Gvec}, {\Gvec}')\, \Cc{\Gvec}\Dc{\Gvec'}
    \,,
}
where $\VL({\Gvec} - {\Gvec'})$
and $\VNL ({\Gvec},{\Gvec'})$
are the matrix elements of local and nonlocal parts of
the pseudopotential, respectively. It is convenient to rewrite
the local part of the pseudopotential and to define the
the following quantities
\begin{subequations}
\begin{align}
\label{V_ei_rho}
    \Veiop
& = \VeiLop + \VeiNLop + N\VL(\mathbf{0})
    \,,
\\
    \VeiLop
& = \Half
    \sum_{\Qvec \neq \mathbf{0}} \VL(\Qvec)
    \left[
        \hat{\rho}({\Qvec}) +
        \hat{\rho}^\dag({\Qvec})
    \right]
    \,,
\\
    \VeiNLop
& = \sum_{\Gvec,\Gvec'}
    \VNL({\Gvec}, {\Gvec'}) \Cc{\Gvec} \Dc{\Gvec'}
    \,,
\end{align}
\end{subequations}
where $N$ is the number of electrons, and the one-body density
operator $\hat{\rho}({\Qvec})$ in this equation is given by
\begin{equation}
    \label{rho}
    \hat{\rho}({\Qvec})
    \equiv
    \sum_{\Gvec,\lambda}
    \Cc{{\Gvec + \Qvec},\lambda} \Dc{{\Gvec},\lambda}
    \,
    \theta
    \left(
        \Ecut - \left|{\Gvec+\Qvec}\right|^2/2
    \right)
    ,
\end{equation}
where the step function ensures that $({\Gvec + \Qvec})$ lies within the
planewave basis, and the summation over electron spins ($\lambda=1,2$)
has been made explicit.

The electron-electron interaction is given by
\eql{Vee}
{
    \Veeop
& = \Half N \xi
\\
& + \,
    \frac{1}{{2\Omega}} {\sum_{ijkl}}'
    \frac{4\pi}{\left| \Gvec_i - \Gvec_k \right|^2} \,
    \delta_{\Gvec_i - \Gvec_k , \Gvec_l - \Gvec_j}^{}
    \delta_{\lambda_i, \lambda_k}^{}
    \delta_{\lambda_j, \lambda_l}^{}
\\
&   \quad \, \times
    \Cc{i} \Cc{j} \Dc{l}\Dc{k}
    \;.
}
The primed summation indicates that the $\Gvec_i = \Gvec_k$
singular term is excluded, due to charge neutrality.
The first term in this equation is a constant due to the
self-interaction of an electron with its periodic images.  
It depends only on the number of electrons in the
simulation cell and the Bravais lattice associated with the periodic
boundary conditions.  The standard Ewald expression for $\xi$ is given by
\cite{Fraser96}
\eql{xi}
{
    \xi
& = \frac{1}{\Omega}
    \sum_{\Gvec \neq \mathbf{0}}
    \frac{\exp(-\pi^2 G^2 / \kappa^2)}{\pi G^2}
  - \frac{\pi}{\kappa^2 \Omega}
    \\
& + \,
    \sum_{\Rvec \neq \mathbf{0}}
    \frac{\erfc(\kappa R)}{R}
  - \frac{2\kappa}{\sqrt{\pi}}
    \,,
}
where $\Rvec$ is a direct lattice vector, and $\xi$ is independent
of the Ewald constant $\kappa$, which only controls the relative
convergence rates of the direct and reciprocal space summations.
For the discussion below, we rewrite the two-body
contribution in Eq.~(\ref{Vee}):
\eql{V_rho-rho}
{
    \Veeop
& = \Half N \xi
  + \frac{1}{2\Omega} \sum_{\Qvec \neq \mathbf{0}}
    \frac{4\pi}{Q^2}\,\hat{\rho}^\dagger(\Qvec) \hat{\rho}(\Qvec)
\\
& - \,
    \frac{1}{2\Omega} \sum_\lambda \sum_{\Gvec,\Gvec'}
    \frac{4\pi}{\left|\Gvec - \Gvec'\right|^2}
    \Cc{\Gvec,\lambda} \Dc{\Gvec,\lambda}
    \,.
}
The third term in Eq. (\ref{V_rho-rho}) is a sum of diagonal one-body
operators arising from the anticommutation of the fermion creation
and destruction operators.

Finally, we can regroup the contributions to the Hamiltonian into constant,
one-body, and two-body parts,
\begin{equation}
    \Hop = H^{(0)} + \Hop^{(1)} + \Hop^{(2)},
\end{equation}
where
\begin{subequations}
\begin{align}
\label{eq:Hterms}
    H^{(0)}
& = \Half N \xi
  + \Vii + N\VL(\mathbf{0})
\\
\begin{split}
    \Hop^{(1)}
& = \Kop + \VeiLop + \VeiNLop
\\
& - \,
    \frac{1}{2\Omega} \sum_\lambda \sum_{\Gvec,\Gvec'}
    \frac{4\pi}{\left|\Gvec - \Gvec'\right|^2}
    \Cc{\Gvec,\lambda} \Dc{\Gvec,\lambda}
\end{split}
\\
    \Hop^{(2)}
& = \frac{1}{2\Omega} \sum_{\Qvec \neq \mathbf{0}}
    \frac{4\pi}{Q^2}\,\hat{\rho}^\dagger(\Qvec) \hat{\rho}(\Qvec)
\end{align}
\end{subequations}

It is convenient to express the two-body part as a quadratic form of
one-body operators (this can always be done, and such forms are not
unique). We use the identity
$\hat{\rho}(-\Qvec) = \hat{\rho}^\dagger(\Qvec)$
to write
%
\begin{equation}
    \hat{H}^{(2)}
  = \sum_{\Qvec \neq \mathbf{0}}
    \frac{\pi}{\Omega Q^2}
    \left[
        \hat{\rho}(\Qvec) \hat{\rho}^\dagger(\Qvec)
      + \hat{\rho}^\dagger(\Qvec) \hat{\rho}(\Qvec)
    \right].
\end{equation}
Defining Hermitian operators $\Aop({\Qvec})$ and $\Bop({\Qvec})$ as
\begin{subequations}
\begin{align}
\label{AB}
    \Aop(\Qvec)
&   \equiv
    \sqrt{\frac{2\pi}{\Omega Q^2}}
    \left[ \hat{\rho}(\Qvec) + \hat{\rho}^\dagger(\Qvec) \right]
    \,,
\\
    \Bop(\Qvec)
&   \equiv
    i \sqrt{\frac{2\pi}{\Omega Q^2}}
    \left[ \hat{\rho}(\Qvec) - \hat{\rho}^\dagger(\Qvec) \right]
    \,,
\end{align}
\end{subequations}
the two-body contribution becomes a simple sum of quadratic operators,
\begin{align}
\label{H2}
    \Hop^{(2)}
& = \frac{1}{4} \sum_{\Qvec \neq \mathbf{0}}
    \left[ \Aop^2({\Qvec}) + \Bop^2({\Qvec}) \right] \notag
\\
& = \frac{1}{2} \sum_{\Qvec > \mathbf{0}}
    \left[ \Aop^2({\Qvec}) + \Bop^2({\Qvec}) \right]
    \,,
\end{align}
where we have used the ${\Qvec} \to -{\Qvec}$ symmetry to obtain the last
expression.

\subsection{Ground state projection and the Hubbard-Stratonovich Transformation}
\label{sec:GS-projection}

The ground state of $\Hop \ket{\PsiGS} = E_0 \ket{\PsiGS}$
is obtained by imaginary-time projection from a trial wave function $\ket{\Psi_T}$
\eql{eq:QMC-proj}
{
    \lim_{n \to \infty}
    {\left(e^{-\Delta \tau (\Hop - E_0)}\right)}^n \ket{\Psi_T}
    =
    \ket{\PsiGS}
    \,,
}
provided $\braket{\Psi_T}{\PsiGS} \neq 0$.  In the present
calculations, $\ket{\Psi_T}$ is a single Slater determinant obtained
from a mean-field calculation.
Expressing the imaginary-time projection in terms of the
small discrete time step $\Delta \tau$ facilitates the separation of
the one- and two-body terms, using the short-time Trotter-Suzuki
decomposition \cite{Trotter1959,Suzuki1976}
\eql{eq:Trotter}
{
    e^{-\Delta \tau \Hop}
& = e^{-(1/2) \Delta \tau \Hop^{(1)}}
    e^{-\Delta \tau \Hop^{(2)}}
    e^{-(1/2)\Delta \tau \Hop^{(1)}}
\\
& + \, O(\Delta \tau^3)
    \,.
}
The application of the one-body propagator
$e^{-(1/2)\Delta \tau \Hop^{(1)}}$
on a Slater determinant $\ket{\phi}$ simply yields another Slater
determinant:
$ \ket{\phi'} = e^{-(1/2) \Delta \tau \Hop^{(1)}} \ket{\phi}$.
The two-body propagator is expressed as an integral of
one-body propagators, using the Hubbard-Stratonovich transformation
\cite{Hubbard,Stratonovich}
\eql{eq:HS-xform0}
{
&   \exp{\left({-\Half \Delta\tau \sum_i\lambda_i \HSop_i^2}\right)}
\\
&\quad\;
    =
    \int \left( \prod_i \frac{d\sigma_i}{\sqrt{2\pi}} \right)
    \exp{\left[\sum_i \left(-\Half \sigma_i^2
    +\sigma_i \sqrt{-\Delta\tau\lambda_i} \, \HSop_i \right)
    \right]}
}
for any one-body operators $\{\,\HSop_i\,\}$.
Thus we have
\eql{eq:HS-xform}
{
    e^{-\Delta\tau\Hop^{(2)}}
  = \left(\frac{1}{\sqrt{2\pi}}\right)^{\!\dim(\bm{\sigma})}
    \!
    \int d\bm{\sigma}  \,
    e^{-(1/2) \bm{\sigma} \cdot \bm{\sigma}}
    e^{\sqrt{\Delta\tau}\, \bm{\sigma} \cdot \hat{\mathbf{v}} }
    \, ,
}
where we introduce a vector $\bm{\sigma} \equiv \{\sigma_i\}$,
whose dimensionality, $\dim(\bm{\sigma})$, is the number of \textit{all\/}
possible $\Qvec$-vectors satisfying $\Qvec = \Gvec - \Gvec'$ for two
arbitrary wave vectors $\Gvec$ and $\Gvec'$ in the planewave basis.
The operators $\hat{\mathbf{v}} \equiv \{ \sqrt{-\lambda_i}\,\HSop_i \}$
are given by the $i\Aop({\Qvec})$ or $i\Bop({\Qvec})$ one-body operators,
since all the $\lambda_i = 1$.

In the original formulations of the
AF QMC method \cite{Blankenbecler1981, Koonin1986},
the many-dimensional integral over the auxiliary
fields $\bm{\sigma}$ in Eq.~(\ref{eq:HS-xform}) is
evaluated by standard Metropolis or heat-bath algorithms.
We instead apply
an importance-sampling transformation \cite{sz-cpmc,sz-hk,ZhangB} to turn the
projection into a branching random walk in an overcomplete
Slater determinant space.
The importance sampling helps guide the random walks according to the projected
overlap with the trial wave function. More importantly, it allows the 
imposition of a constraint to control the phase problem.


A phase problem arises 
for a general repulsive two-body interaction, because 
the $\lambda_i$ cannot be made all negative. In other words,
not all components of 
the operator $\hat{\mathbf{v}}$ can be made real.
 (Although this is in principle possible by
an overall shift to the potential \cite{Koonin1986} or by introducing
many more auxiliary fields,
they both
cause large fluctuations \cite{sz-hk,Silvestrelli93}.)
As the random proceeds, the projection in Eq.~(\ref{eq:HS-xform})
\begin{equation}
\label{randwlk-stp}
    \ket{\phi'}
    \leftarrow
    \exp(\sqrt{\Delta\tau}\, \bm{\sigma} \cdot \hat{\mathbf{v}}) \ket{\phi}
\end{equation}
by a complex $\hat{\mathbf{v}}$
causes the orbitals in the Slater determinants $\left|\phi\right\rangle$ to become
complex.  For large imaginary projection times, the
phase of each $\left|\phi\right\rangle$ becomes random, and the stochastic
representation of the ground state $\ket{\PsiGS}$ becomes dominated by
noise. This leads to the phase problem and the divergence of the
fluctuations. The phase problem is of the same origin as the sign
problem that occurs when the one-body operators $\hat{\mathbf{v}}$ are real, but
is more severe because, instead of a $+\left|\phi\right\rangle$ and
$-\left|\phi\right\rangle$ symmetry \cite{kalos91,sz-cpmc}, there is now an
infinite set $\{ e^{i\theta} \left|\phi\right\rangle, \theta \in
[0,2\pi) \}$, among which the Monte Carlo sampling cannot distinguish.

The phaseless AF QMC method \cite{sz-hk} used in this paper controls
the phase/sign problem in an approximate manner using a trial wave
function. The method uses a {\em complex} importance function, the
overlap $\langle \Psi_T|\phi\rangle$, to construct phaseless
random walkers, $|\phi\rangle/\langle \Psi_T|\phi\rangle$,
which are invariant under a phase gauge transformation. The resulting
two-dimensional diffusion process in the complex plane of the overlap
$\langle \Psi_T|\phi\rangle$ is then approximated as a
diffusion process in one dimension.
Additional implementation details can be found in 
Refs.~\onlinecite{sz-hk,Al-Saidi,Purwanto2004}.
The phaseless constraint is
different from the nodal condition imposed in fixed-node DMC, since the
phaseless constraint confines the random walk in Slater determinant
space according to its overlap with a trial wave function, which is a
global property of $\left|\phi\right\rangle$.  Thus, the phaseless
approximation can behave differently from the fixed node approximation
in DMC.

Finally, we describe the use of fast Fourier transform (FFT) with a planewave basis to
efficiently implement the random walk projection given by
Eqs.~(\ref{AB}), (\ref{H2}), (\ref{eq:HS-xform}), and (\ref{randwlk-stp}).
For example,
\eql{eq:proj-FFT}
{
&\exp \left(
     \sum_{\Qvec} \sqrt{\frac{\Delta\tau}{Q^2}} \,
                  \sigma(\Qvec) \hat{\rho}(\Qvec)
      \right) \ket{\phi} \\
&\qquad \simeq
 \sum_{n = 0}^{n_{\max } } \frac{1}{n!}
     {\left ( \sqrt { \frac{\Delta \tau} {Q^2} }  \,
             \sigma(\Qvec)\hat{\rho}(\Qvec) \right)\!}^n \ket{\phi}
 \, .
}
Terms in the series can be evaluated as an iterative FFT, since 
$\rho (\bf{Q}) \ket{\phi}$ is just a convolution.
For typical values of $\Delta \tau$, we find that $n_{\max} \simeq 4$ accurately reproduces
the propagator.

\subsection{Ground-state mixed estimator}
\label{sec:mixed-est}

The ground state energy $E_0$ can then be obtained by the mixed estimator 
\begin{equation} 
\label{E_estimat}
    E_0 = \frac{\ME{\Psi_T}{\Hop}{\PsiGS}}
               {\braket{\Psi_T}{\PsiGS}}
        = \lim_{\beta \to \infty}
          \frac{\ME{\Psi_T}{\Hop e^{-\beta\Hop}}{\Psi_T}}
               {\ME{\Psi_T}{e^{-\beta\Hop}}{\Psi_T}}
    \,,
\end{equation}
which is evaluated periodically from the ensemble of Slater determinants generated in the
course of the random walks.
In the phaseless AF QMC method, an importance sampling transformation \cite{sz-hk}
leads to a stochastic representation of the ground-state wave function in
the form of
\begin{equation}
    \ket{\PsiGS}
  = \sum_\phi w_\phi
    \frac{\ket{\phi}}{\braket{\Psi_T}{\phi}}
    \,.
\label{eq:wf_MC_imp}
\end{equation}
This means the mixed estimate for the energy is given by
\begin{equation}
    E_0^{\textrm{MC}}
  = \frac{\sum_\phi w_\phi E_L[\phi]}{\sum_\phi w_\phi}
    \,,
\label{eq:mixed_w_EL}
\end{equation}
where the local energy is defined as
\begin{equation}
    E_L[\phi]
    \equiv
    \frac{\ME{\Psi_T}{\Hop}{\phi}}
         {\braket{\Psi_T}{\phi}}
    \,.
\label{eq:El}
\end{equation}
%
%

Matrix elements of one-body terms in the local energy (and other similar 
estimators) can be
expressed in terms of the one-body Green's functions\cite{sz-cpmc,ZhangB}
\begin{equation} 
\label{G1}
    G_{ji}
  = \langle {c_j^\dag c_i } \rangle
    \equiv
    \frac{\ME{\Psi_T}{\Cc{j} \Dc{i}}{\phi}}
         {\braket{\Psi_T}{\phi}}
    \,.
\end{equation}
The Green's function can be expressed in terms of the one-particle
orbitals in the $\ket{\Psi_T}$ and $\ket{\phi}$ Slater determinants
as follows. A general Slater determinant $\ket{\phi}$ can be written as
\begin{equation} 
\label{phi}
| \phi \rangle \equiv \phi_1^\dag \phi_2^\dag \cdot \cdot
\cdot \phi_N^\dag | 0 \rangle 
\end{equation}
where the $\phi_i^\dag$ creates an electron in the orbital $i$
\begin{equation} 
\label{phi2}
\phi_i^\dag \equiv \sum\limits_{j} c_j^\dag \Phi_{ji},
\end{equation}
and $j$ labels 
the one-particle orthogonal basis functions, which are planewaves in the present case. The
$\Phi_{ji}$ are the elements of a $M \times N$ dimensional matrix $\bm
\Phi$. Each column of the matrix $\bm{\Phi}$ represents a single-particle
orbital expressed as a sum of planewaves. It is a well-known
result that the overlap of two Slater determinants is given by the determinant of
the overlap matrix of their one particle orbitals
\begin{equation} 
\label{overlap}
\left \langle \Psi_T | \phi \right \rangle =
{\rm det} \left ( {\bm \Psi}_T^\dag {\bm \Phi}   \right ).
\end{equation}
Finally, it can be shown that the Green's function can be expressed
as\cite{Loh1992}
\begin{equation} 
\label{G1_a}
G_{ji} = 
\left [ {\bm \Phi} 
\left ( {\bm \Psi}_T^\dag {\bm \Phi} \right )^{-1}
{\bm \Psi}_T^\dag
\right ]_{ij}
\end{equation}

Hamiltonian matrix elements of two-body terms in the mixed estimator
are expressed in terms of the two-body Green's function, which 
can be written as products of one-body Green's functions
using the Fermion anticommutation properties,
\begin{align}
\begin{split}
\label{G2G1G1}
    \langle {\Cc{i} \Cc{m} \Dc{n} \Dc{j}} \rangle
&   \equiv
    \frac{\ME{\Psi_T}{\Cc{i} \Cc{m} \Dc{n} \Dc{j}}{\phi}}
         {\braket{\Psi_T}{\phi}}
\\
& = G_{ji} G_{nm} - G_{ni} G_{jm} \; .
\end{split}
\end{align}
%
Rather than directly implementing Eq.~(\ref{G2G1G1}), it is more
efficient to use fast Fourier
transformations to take advantage of locality in real space.  
The computer time to calculate the mixed estimator
then scales as $N^2 M \log(M)$, where N is the number of electrons and
M is the number of planewaves.

\subsection{Trial wave function}
\label{sec:psiT}

The trial wave function $\ket{\Psi_T}$ determines the systematic accuracy of our calculations, 
due to the use of the phaseless approximation. Its quality also affects
the statistical precision.
We use a single Slater determinant as the
trial wave function, which is obtained either from a
DFT calculation or HF calculation.  The DFT wave functions were generated
self-consistently with \ABINIT{} \cite{abinit},
using a planewave basis and the local density approximation (LDA). The HF
wave functions were obtained from an in-house planewave-based Hartree-Fock
program. 
In both cases, identical setup is used in the independent-electron
calculation as in the corresponding AF QMC calculations.

\subsection{Pseudopotentials} 
\label{sec:pseudos}

Norm-conserving
pseudopotentials \cite{KB} are used in the
present calculations. 
Pseudopotentials are necessary to
keep the basis size tractable by
eliminating the highly-localized core states.
Pseudopotential transferability
is a source of potential errors, however, especially since the 
pesudopotentials used here are generated from independent-electron 
calculations. Such pseudopotentials are quite routinely employed in QMC 
and other many-body calculations and have proved very useful. But their 
transferability is not nearly as extensively quantified and studied as 
in standard independent-electron calculations.
Thus one of our goals here is to examine the 
use of such pseudopotentials in the many-body AF QMC framework.

The pseudopotential has been adapted to take the Kleinman-Bylander (KB)  \cite{KB}
form suitable for planewave calculations,
\begin{subequations}
\begin{align}
\label{eq:PSP-KB}
    \Veiop^{\textrm{(KB)}}(\mathbf{r})
& = \VeiLop^{\textrm{(KB)}}(\mathbf{r})
  + \VeiNLop^{\textrm{(KB)}}(\mathbf{r})
    \,,
\\
    \VeiNLop^{\textrm{(KB)}}(\mathbf{r})
& = 
    \sum_{l,m}
    \frac{\ket{V_l \varphi_l Y_{lm}} \bra{V_l \varphi_l Y_{lm}}}
         {\ME{\varphi_l}{V_l}{\varphi_l}}
    \,,
\end{align}
\end{subequations}
where $\varphi_l$ is the pseudo\-orbital for the $l$-th angular momentum
component.
In our case, we use the neutral atomic reference state (with an LDA-type
Hamiltonian) to generate the pseudo\-orbitals.

To examine the effects of pseudopotentials on the accuracy of the AF QMC calculations,
we employ two pseudopotentials in this study:
the optimized LDA-based pseudopotential~\cite{Rappe} generated using the OPIUM
\cite{opium} package, and
the HF-based effective core potential developed by Ovcharenko, Aspuru-Guzik, and Lester
\cite{LesterECP2, LesterECP1}.
We will subsequently refer to these pseudopotentials as
OPIUM and \ECP{}, respectively. The semilocal \ECP{} pseudopotentials
were converted to the fully nonlocal KB form, using
the atomic LDA ground state wave function as the reference state.
The \ECPpsp{} is not used in molecules, because it lacks a $d$ projector.
An illustration of this point is given in Table~\ref{tab:PSP-quality}.

Table~\ref{Ecut+Psp} 
gives parameters describing the OPIUM pseudopotentials.
The second column shows the cutoff energy for each atomic species.
The same cutoff energies are also used in our calculations
with \ECPpsp{s}. 
(The parameters of \ECPpsp{s} have been published in
Ref.~\onlinecite{LesterECP2}.)
These were tested for convergence with LDA and then verified with AF
QMC calculations.

\begin{table}[!htbp]

\caption{\label{Ecut+Psp}
Optimized $\Ecut$ and OPIUM pseudopotential parameters used in the
calculations.
Each angular component ($l$) of the pseudopotential has its own cutoff
radius ($r_c$).
}
\begin{ruledtabular}
\begin{tabular}{cccccl}
        &        & \multicolumn{3}{c}{$r_c$ (units of $\bohr$)} &  Reference \\
\cline{3-5}
Species & $\Ecut$ (Ha) &   $l = 0$   &   $l = 1$   &   $l = 2$  &  configuration \\
\hline
   Al   & \ \ $7.50$   &             &             &            &   \\
   Si   & \ \ $6.13$   &    $2.20$   &    $2.20$   &    $2.50$  &  [Ne] $3s^2
3p^2$ \\
   P    &    $18.00$   &    $1.75$   &    $1.75$   &    $2.50$  &  [Ne] $3s^2 3p^{2.5} 3d^{0.5}$ \\
   S    &    $19.00$   &    $1.75$   &    $1.75$   &    $1.75$  &  [Ne] $3s^2 3p^{3.5} 3d^{0.5}$ \\
   Cl   &    $18.00$   &    $1.75$   &    $1.75$   &    $2.50$  &  [Ne] $3s^2 3p^{4.5} 3d^{0.5}$ \\
   As   &    $18.00$   &    $1.80$   &    $1.80$   &    $2.50$  &  [Ar] $4s^2 4p^{2.5} 4d^0$ \\
\end{tabular}
\end{ruledtabular}
\end{table}

\section{Convergence studies}
\label{sec:convergence}

To achieve high accuracy and to minimize the computational cost, one
should optimize the calculations with respect to
the number of basis functions, the
supercell size, and the magnitude of the Trotter time step.
In this section, we illustrate 
the convergence of our method with respect to these
parameters. 

Ionization energies are defined as
$\IP \equiv E(N-1) - E(N)$ and $\IIP \equiv E(N-2) - E(N)$,
for the singly- and doubly-ionized atoms, respectively,
where $N$ is the number of electrons in the neutral atom. 
The dissociation energy $D_e$ is calculated as the
difference between the total energy of the dimer at the
experimental equilibrium bond length and the energy of the isolated atoms,
$\De \equiv 2E_{\textrm{atom}} - E_{\textrm{dimer}}$.

\subsection{Planewave convergence}
\label{sec:PWconverg}

Convergence with respect to the planewave cutoff energy $\Ecut$ depends on
both $\Hop^{(1)}$ and $\Hop^{(2)}$ in Eq.~(\ref{eq:Hterms}).  The $\Hop^{(1)}$
dependence is similar to that in independent-electron calculations. 
Convergence requires that $\Ecut$ is sufficient for the 
``hardness'' of the pseudopotential and the electronic density variations. 
The $\Hop^{(2)}$ dependence has to do with the scattering matrix elements in
the two-body interaction. In the uniform electron gas, for example, 
$\Hop^{(1)}$
requires an $\Ecut$ given by the Fermi energy $E_{\rm F}$ 
(for restricted HF), 
while $\Hop^{(2)}$ will lead to a
finite convergence error for any 
finite $\Ecut$, 
which decreases as $\Ecut$ is increased and, for a 
fixed $\Ecut/E_{\rm F}$, becomes more pronounced as 
the electronic density is decreased.

\begin{figure}[!hbtp]  
\includegraphics[scale=.33]{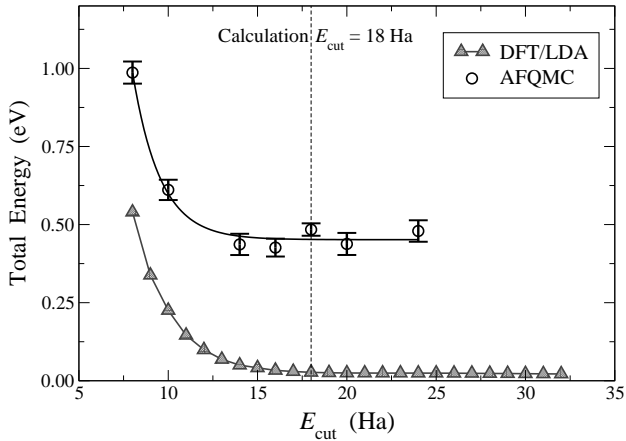}
\caption{\label{ecut} Convergence of the total energy of a phosphorus atom
in a $14 \times 14 \times 14 \bohr^3$ unit cell.
The OPIUM pseudopotential is used here.
The convergence behavior is similar in both LDA and AF QMC methods.
A constant shift is added to each data set for convenience,
such that the converged energies
are approximately $0$ eV and $0.5$ eV for LDA and AF QMC, respectively.
}
\end{figure}

Fig.~\ref{ecut} shows the
phosphorus atom total energy as a function of the planewave cutoff energy
$\Ecut$ for both AF QMC and LDA. 
The calculations were done for a fixed supercell size and
a pseudopotential whose design cutoff energy is 18 Ha. In LDA the total energy
was converged to within 5\,meV at this cutoff.
The energy decreases monotonically with increasing $\Ecut$
in both calculations. We see that the AF QMC convergence behavior
is similar to LDA, indicating that 
the AF QMC convergence error 
from $\Hop^{(2)}$ is much smaller here than that from  $\Hop^{(1)}$.
This trend was found to be typical of the systems studied in this paper 
with the chosen pseudopotentials.
Table~\ref{Ecut+Psp} shows the cutoff energy for each atomic species.
In subsequent calculations for phosphorus, for example,
we used $\Ecut = 18$\,Ha
in the AF QMC, as indicated by the vertical dashed line in Fig.~\ref{ecut}.

\subsection{Supercell size convergence}
\label{sec:Supercellconverg}

\begin{figure}[!hbtp]  
\includegraphics[scale=.33]{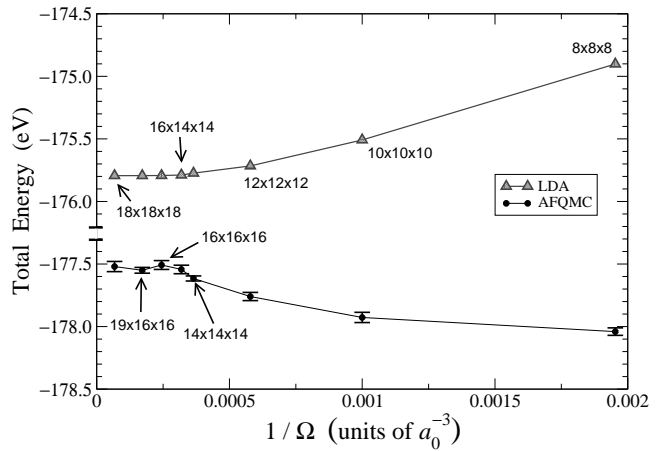}
\caption{\label{PE} Convergence of the phosphorus
total energy with respect to the inverse of the simulation cell volume
$\Omega$.
The triangles denote the results of LDA calculations
while the solid circles denote those of the AF QMC.
The OPIUM pseudopotential is used here.
}
\end{figure}

\begin{figure}[!hbtp]  
\includegraphics[scale=.72]{\FIGDIR{}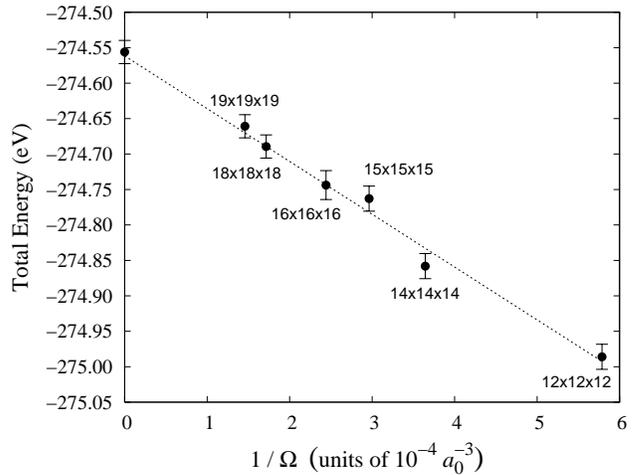}
\caption{\label{fig:S-Etotal-1/V}
Convergence of the AF QMC total energy with simulation cell size
for Sulfur, plotted as a function of the inverse volume, from box sizes
$12 \times 12 \times 12 \bohr^3$ through $19 \times 19 \times 19 \bohr^3$.
The \ECPpsp{} is used here.
The dotted line shows the fitting of the total energy as
a linear function of $1 / \Omega$.
The data point shown at $1 / \Omega = 0$ shows the extrapolated energy at
infinite box size.  
}
\end{figure}

Due to the periodic boundary conditions imposed in the calculations, the
interactions between electrons in the simulation cell and their
periodic images give rise to finite-size errors. 
To study the behavior of these errors, a series of
LDA and AF QMC calculations were performed using different system
sizes for cubic (and some tetragonal) shaped cells. Results for the phosphorous atom
are shown in Fig.~\ref{PE} as a function of supercell size.  The AF QMC energies are seen to converge from below while
the LDA energy converge from above.
This is not surprising, since LDA treats the supercell Coulomb interaction differently 
from a many-body approach such as AF QMC. 
\COMMENTED{
Fig.~\ref{PE}
shows the finite-size effects in the two
methods.
}
Fig.~\ref{fig:S-Etotal-1/V} shows that the total energy from AF QMC for the sulfur atom 
is nearly a linear function of $1/\Omega$ for this range of supercell sizes.
At the largest $19 \times 19 \times 19 \bohr^3$ supercell size,
the total energy is converged to within about $\sim 0.1$ eV.
\COMMENTED{
The remaining finite-size error, estimated via $1/V$ extrapolation, is
about $0.1$ eV.
This trend, however, is much weaker for the ionization energies, where it
has been very much converged for simulation boxes equal to or greater than
$16 \times 16 \times 16 \bohr^3$.
}

\begin{figure}[!hbtp]  
\includegraphics[scale=.33]{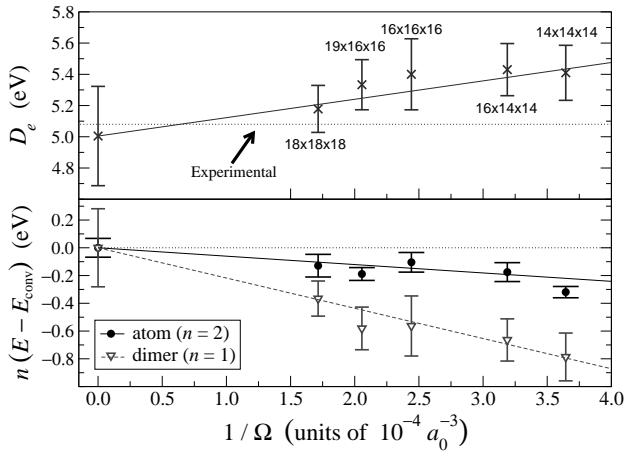}
\caption{\label{DE_E_2E} Phosphorus dissociation energy (top panel)
 and the total energies of P (times two) and P$_2$ (bottom panel) for
 different supercell sizes,
 computed using the OPIUM pseudopotential.
 The dissociation energy obtained with the
 largest simulation cell, $18 \times 18 \times 18 \bohr^3$ is within
 0.1 eV of the experimental value of 5.08 eV (indicated by the horizontal
 dotted line in the top panel).
 $\Econv$ is the energy extrapolated to $1/\Omega \to 0$
 (using cubic cells only).
 }
\end{figure}

To demonstrate the supercell size effect on the energy differences,
Fig.~\ref{DE_E_2E} shows the calculated dissociation 
energy of P$_2$. 
The top panel shows the supercell
size dependence of the dissociation energy, and the bottom panel
illustrates the convergence error of P and P$_2$ energies for supercells ranging
from $14 \times 14 \times 14 \bohr^3$ through $18 \times 18 \times 18 \bohr^3$.
The total atom energy from different supercells deviates no more than $0.2$ eV 
from that of
the largest supercell. The dimer total energy, on the other hand,
shows stronger finite-size effect. Most of the finite-size
error in the dissociation energy thus arises from the dimer energy. 
To avoid any irregular dependence of the energy on the aspect ratio
(cubic vs. tetragonal supercells), 
only the cubic supercells were used in the extrapolation.
\COMMENTED{
According to Fig.~\ref{DE_E_2E},
phosphorus dimers for $14 \times 14 \times 14 \bohr^3$ reaches about $0.4$
eV from $\Econv$, indicating that most of the finite-size
error in the dissociation energy arises from the dimer energy.
}

\begin{figure}[!hbtp]  
\includegraphics[scale=.346]{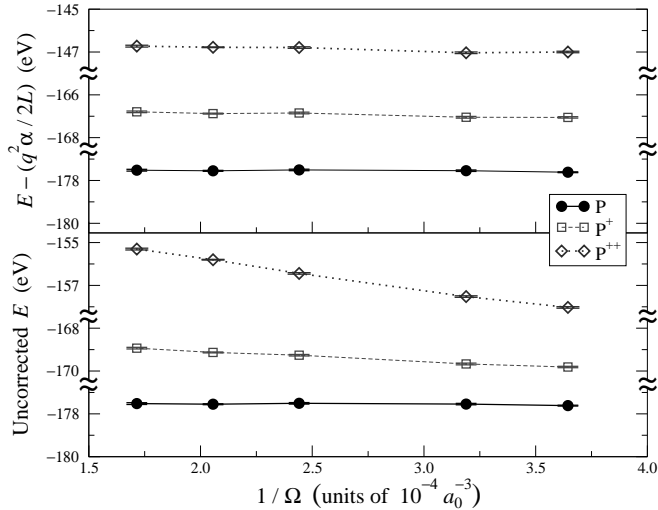}
\caption {\label{PECorr}
An illustration of the size effect in the calculation of charged
atoms.
The top panel shows the corrected total energies for P, P$^+$ and
P$^{++}$, while the bottom panel shows the uncorrected ones.
Cell sizes used in this study are the same as in Fig.~\ref{DE_E_2E}.
The errorbars are smaller than the point size.
}
\COMMENTED{NOTE:
The extra correction (-q * xi) has been included in the uncorrected
QMC total energies above!
That correction was necessary due to the unfortunate convention
inside our QMC code.
}
\end{figure}

In the ionization energy calculations, the supercells are charged
$+|e|$ and $+|2e|$ for the $X^+$ and $X^{++}$ species, respectively.
Charged supercells are ill-defined under periodic boundary conditions,
so an additional neutralizing background charge is introduced
to maintain charge neutrality \cite{MakovPayne1,MakovPayne2}.
As discussed by Makov and Payne \cite{MakovPayne1,MakovPayne2}, 
a leading behavior, $q^2\alpha /2L$, arises from the self-interaction of the neutralizing charge
with its periodic images,
where $\alpha$ is the (supercell-dependent)
Madelung constant, $q$ is neutralizing charge, and $L^3 = \Omega$.
Correction of the total energy by the
leading term leads to more rapid size convergence.
Figure~\ref{PECorr} illustrates this effect.
The bottom panel shows the slow convergence of the total energy 
with the system size in charged systems, while the upper panel shows the
more rapid convergence after the correction has been made, i.e., the slowly convergent
$q^2\alpha /2L$ contribution has been subtracted.

\subsection{Trotter time step error}
\label{sec:Trotter}

\begin{figure}[!hbtp]  
\includegraphics[scale=.33]{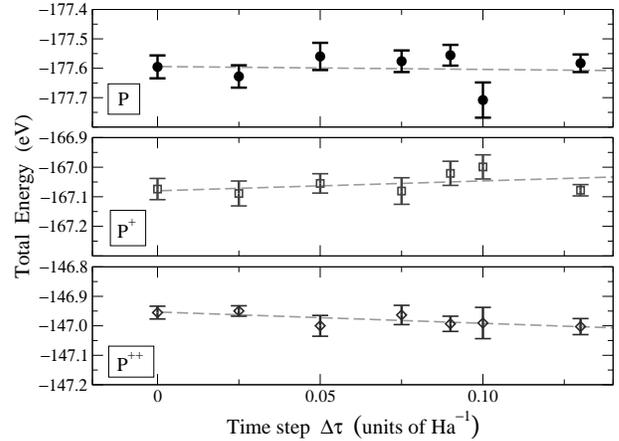}
\caption{\label{trotter_P}
Trotter errors for P, P$^+$ and P$^{++}$ energies (shown in the top,
middle, and bottom panels, respectively) in a
$14 \times 14 \times 14 \bohr^3$ simulation box.
The data points at $\Delta \tau = 0$ represent the Trotter extrapolation.
}
\end{figure}

The Trotter error arises from neglecting higher order terms of
the imaginary-time propagator, $e^{-\Delta\tau{\Hop}}$, when we
apply the Trotter-Suzuki decomposition in Eq.~(\ref{eq:Trotter}).
\COMMENTED{
It is important to choose an appropriate $\Delta \tau$ value.
If $\Delta \tau$ is too small, the calculations will be computationally
expensive.
On the other hand, a large $\Delta \tau$ causes a large Trotter error.
}
The Trotter error can be eliminated by extrapolation,
as demonstrated in Fig.~\ref{trotter_P} for 
P, P$^+$, and P$^{++}$.
The results reported in the next section use either a linear extrapolation 
a fixed $\Delta \tau$ ($\sim 0.05$ Ha$^{-1}$) which is
sufficiently small so that the Trotter error 
is well within the statistical error.

\section{Results}
\label{sec:results}



Table \ref{tab:DE} shows dissociation energies
from planewave AF QMC calculations compared with
experimental values and with LDA, GGA, HF, and diffusion Monte
Carlo (DMC) calculated results.
LDA trial wave functions were used in the molecular calculations,
corresponding to the following electronic 
configurations: $\sigma ^2 _{\rm{3s}}
\sigma ^{*2} _{\rm{3s}}\sigma ^2 _{\rm{3p}}\pi ^4 _{\rm{3p}}$ for
P$_2$, $\sigma ^2 _{\rm{3s}} \sigma ^{*2} _{\rm{3s}}\sigma ^2
_{\rm{3p}}\pi ^4 _{\rm{3p}}\pi ^{*2} _{\rm{3p}}$ for S$_2$, $\sigma ^2
_{\rm{3s}} \sigma ^{*2} _{\rm{3s}}\sigma ^2 _{\rm{3p}}\pi ^4
_{\rm{3p}}\pi ^{*4} _{\rm{3p}}$ for Cl$_2$, and $\sigma ^2 _{\rm{4s}}
\sigma ^{*2} _{\rm{4s}}\sigma ^2 _{\rm{4p}}\pi ^4 _{\rm{4p}}$ for
As$_2$. All dimers except S$_2$ have ``closed-shell''
configurations in which all occupied orbitals are fully filled. All the calculations
used OPIUM pseudopotentials.
\COMMENTED{
The periodic supercell sizes for P$_2$, S$_2$, Cl$_2$ and As$_2$ are
(in units of $\bohr^3$)
$19 \times 16 \times 16$, $16 \times 14 \times 14$, $18 \times 18 \times 18$
and $20 \times 16.5 \times 16.5$, respectively.
}
\begin{table}[!htbp]
\caption{\label{tab:DE} Calculated dissociation energies (in eV) 
using LDA, AF QMC, DMC, HF, and GGA methods.
Experimental values are in the last column (with
the zero-point energy
removed).
The AF QMC calculations used LDA trial wave functions, except for
sulfur, where we have also used unrestricted HF trial wave functions.
Values from the largest supercell are shown here, since convergence has
been reached.
The statistical errors are given in parentheses.
The HF results are obtained from an in-house planewave based code. 
The AF QMC result for Si$_2$ is taken from Ref.~\onlinecite{sz-hk}.
}
\begin{ruledtabular}
\begin{tabular}{ccccccc}
       & LDA  & AF QMC (LDA) & DMC \cite{GrossmanDMC}
                                     & HF
                                            & GGA \cite{PPP}
                                                   & Expt. \cite{HBconst} \\

\hline
Si$_2$ & 3.88  & 3.12(8)$\,\,$ &     &      &      & 3.21(13) \\
P$_2$  & 5.97  & 5.19(16)  & 4.73(1) & 1.74 & 5.22 & 5.08  \\

S$_2$  & 5.61  & 4.63(17)  & 4.31(1) & 2.29 & 4.94 & 4.41 \\
       &   & \,\,4.48(19)\footnotemark& & &  &       \\

Cl$_2$ & 3.12  & 2.78(10)  & 2.38(1) & 0.67 & 2.76 & 2.51  \\

As$_2$ & 5.04  & 3.97(17)  &         &      &      & 3.96 \\

\end{tabular}

\end{ruledtabular}

\footnotetext{Using HF trial wave function.}

\end{table}

Both restricted 
and unrestricted trial wave functions were
tested in the AF QMC calculations, but 
there were no difference within statistical errors.
(In restricted trial wave functions, the orbitals with minority spin are
identical to the corresponding majority-spin orbitals.)
Unrestricted HF trial wave functions were also used to calculate the
dissociation energy of S$_2$, which
has an
open-shell configuration.
As shown in Table~\ref{tab:DE}, there is no difference within statistical errors.

The overall agreement between the AF QMC results and experiment is
very good. 
The LDA and GGA slightly
overestimate the dissociation energy, while the HF method significantly
underestimates it. 
The heaviest dimer we calculated, As$_2$, 
is also in  excellent agreement with experiment and compares favorably
with results from other
quantum chemistry methods \cite{SakaiMiyoshi,Moshizuki}. 

\begin{table*}
\newcommand\blank[1] {\multicolumn{#1}{c}{}}
\caption{\label{tab:IP+IIP}
First and second ionization energies for several atoms, in eV.
We employ two pseudopotentials, the LDA-based OPIUM pseudopotential and
HF-based effective core potential (\ECP{}).
In most cases, we use LDA trial wave functions (see text).
The LDA calculations were done using \ABINIT{}.
The HF results were obtained using an in-house HF program
for OPIUM pseudopotential and
\GAUSSIAN[]\cite{Gau98,Gau03} for \ECP{}.
Experimental values are taken from Ref.~\onlinecite{HBconst}.
The quantity $\Delta\textrm{IP}$ below is the difference
between the AF QMC-calculated and the experimental ionization energies,
for which we show the average and r.m.s. average over all the species
shown here.
}
\setlength\tabcolsep{5pt}
\begin{tabular}{c l   r@{.}l  r@{.}l  r@{.}l  r@{.}l   c   r@{.}l  r@{.}l  r@{.}l  r@{.}l}
\hline
\hline
 &
 &
\multicolumn{8}{c}{IP ($X \rightarrow X^+$)} &&
\multicolumn{8}{c}{IIP ($X \rightarrow X^{++}$)}\\
\cline{3-10}
\cline{12-19}
 &
Pseudo &
\multicolumn{2}{c}{HF} &
\multicolumn{2}{c}{LDA} &
\multicolumn{2}{c}{AF QMC} &
\multicolumn{2}{c}{Expt} &\quad&
\multicolumn{2}{c}{HF} &
\multicolumn{2}{c}{LDA} &
\multicolumn{2}{c}{AF QMC} &
\multicolumn{2}{c}{Expt} \\
\hline
\vspace{0.15cm}
Al & \ECP{} &\blank{2}&  5&88   &  5&88(2)  &  5&99   &&\blank{2}&  24&46  &  24&66(2)  &  24&81  \\
\vspace{0.15cm}
Si & OPIUM  &\blank{2}&\blank{2}&  8&18(2)  &  8&15   &&\blank{2}&\blank{2}&  24&59(4)  &  24&50  \\
P  & OPIUM  &  9&97   & 10&57   & 10&74(6)  & 10&49   &&  29&41  &  30&42  &  30&79(6)  &  30&26  \\
\vspace{0.15cm}
   & \ECP{} &  9&94   & 10&41   & 10&61(3)  &\blank{2}&&  29&11  &  29&97  &  30&28(6)  &\blank{2}\\
S  & OPIUM  &  9&33   & 10&45   & 10&09(7)  & 10&36   &&  32&42  &  33&85  &  34&16(7)  &  33&67  \\
\vspace{0.15cm}
   & \ECP{} &  9&21   & 10&35   & 10&08(2)  &\blank{2}&&  32&06  &  33&51  &  33&61(2)  &\blank{2}\\
Cl & OPIUM  & 11&69   & 13&12   & 12&96(11) & 12&97   &&  34&36  &  36&92  &  36&76(10) &  36&78  \\
\vspace{0.15cm}
   & \ECP{} & 11&76   & 13&02   & 12&89(6)  &\blank{2}&&  34&30  &  36&64  &  36&25(6)  &\blank{2}\\
\hline
\multicolumn{4}{l}{$\langle\Delta\textrm{IP}(\textrm{OPIUM})\rangle$}
                      &\blank{2}&  0&00     &\blank{2}&&\blank{2}&\blank{2}&   0&27     &\blank{2}\\
\multicolumn{4}{l}{$\langle\Delta\textrm{IP}(\textrm{OPIUM})\rangle_{\mathrm{rms}}$}
                      &\blank{2}&  0&19     &\blank{2}&&\blank{2}&\blank{2}&   0&36     &\blank{2}\vspace{0.15cm}\\
\multicolumn{4}{l}{$\langle\Delta\textrm{IP}(\textrm{\ECP{}})\rangle$}
                      &\blank{2}&--0&09     &\blank{2}&&\blank{2}&\blank{2}& --0&18     &\blank{2}\\
\multicolumn{4}{l}{$\langle\Delta\textrm{IP}(\textrm{\ECP{}})\rangle_{\mathrm{rms}}$}
                      &\blank{2}&  0&16     &\blank{2}&&\blank{2}&\blank{2}&   0&28     &\blank{2}\\
\hline
\hline
\end{tabular}

\end{table*}


We show in Table~\ref{tab:IP+IIP} the first and second ionization energies
of Al, Si, P, S and Cl. For comparison, experimental values and results from
LDA and HF calculations are also shown.
The AF QMC calculations are performed using 
both the OPIUM and \ECP{} pseudopotentials together with
LDA trial wave functions, except for 
sulfur with the  \ECP{} pseudopotential, where HF trial wave functions are
also used.
Again, no statistically significant dependence on the trial wave function is seen.
%
The tabulated results are converged with respect to size effects,
which are negligible compared to the
QMC statistical error for supercells larger than
$16 \times 16 \times 16 \bohr^3$.

Ionization energies obtained using the LDA are generally in very good agreement with
experiment, and
the present results are consistent with this, with deviations 
typically within $0.1$ eV for IP and $0.3$ eV for IIP, regardless of 
which pseudopotential is used.
The HF method results in larger deviations of 
$0.5$--$1.3$ eV compared to the experiment.

AF QMC results for Al and Si are in good agreement 
with experiment, and 
for Si the agreement is comparable to that
obtained using DMC\cite{LeeNeeds},   
$8.166(14)$ eV and $24.444(22)$ eV for IP and IIP, respectively.
For P, S, and Cl, however, the agreement between AF QMC and
experiment is not as uniform.
In particular, there is a significant dependence on the choice of the
pseudopotential.
For P and S, the AF QMC ionization energies are better
estimated using \ECP{}, while for Cl, OPIUM pseudopotential gives better results.
We will discuss this dependence in the next section.
Agreement between the best AF QMC and experiment values is
in general very good.

\section{Discussion}
\label{sec:discussion}

For dissociation energies, the agreement between AF QMC and experiment was uniformly very good.
The
appearance of larger discrepancies between AF QMC and experiment
for ionization energies
is somewhat surprising. 
The AF QMC calculations above were systematically converged with respect
to finite size effects, Trotter time step error, and planewave basis size.
Two remaining possibilities, errors arising from the use of the 
phaseless approximation and errors due to the use of pseudopotentials, are discussed
in this section. 

Our overall experience with the phaseless AF QMC\cite{sz-hk, AFQMC-CPC2005,
Al-Saidi_TMO, Al-Saidi, Purwanto2004} suggests that the error due to the
phaseless approximation itself is typically small.
Recent AF QMC calculations using a Gaussian basis \cite{Al-Saidi}
show that, in a variety of atoms and molecules, the AF QMC agrees well with experiment or high-level
quantum chemistry methods such as 
the coupled cluster
 with single and double excitations and
  perturbative corrections for triple excitations [CCSD(T)]\cite{CCSD_Cizek,
CCSD_Cizek_ACP, CCSD_Purvis, CCSD_T_pertb}.
Open-shell systems such as P$^+$, P$^{++}$, S, S$^{++}$, Cl, and Cl$^+$, 
where the $p$ shell
is neither half- nor fully-filled, tend to be more difficult to treat
in general \cite{OpenShell}.
In these cases a single-determinant trial wave function breaks the
symmetry by using only one of the degenerate states, 
and the phaseless approximation could
affect the accuracy of the results.
Some indication of this may be present in our results (see Table~\ref{tab:IP+IIP})
where larger errors 
are observed for energy differences between half-filled and
``open-shell'' systems, such as the P$^+$ ionization energy.
It is possible to use multideterminant trial wave functions in these
cases.
We have done several tests, in which we ``symmetrize'' the trial
wave function, resulting in a linear combinations of three different
determinants.
This is designed to equally treat the $p_x$, $p_y$, and $p_z$ orbitals in
the open shell.
However, there seems to be no observable improvement over the
single-determinant trial wave function at the level of statistical accuracy in
this paper.

To isolate and quantitatively evaluate the errors due to the phaseless approximation
on the second-row atoms studied here, 
we performed
calculations with both AF QMC and CCSD(T) methods for Cl, Cl$^+$, and Cl$^{++}$,
using identical Gaussian basis sets and the \ECPpsp{}. 
(The \ECPpsp{} was chosen since its form is already compatible
with standard quantum chemistry programs.)
Thus, both the AF QMC and CCSD(T) methods are applied to the same many-body 
Hamiltonian, expressed in the Hilbert space spanned by the selected Gaussian basis set.
The CCSD(T) method is approximate but is known to be very accurate 
in atoms and in molecules near equilibrium. 
For the comparison, we employed an uncontracted
aug-cc-pVDZ\cite{aug-cc-pVXZ, PNL-basisform} basis set, where Gaussian
functions
with exponents larger than $98$ are removed, resulting in a $(7s7p2d)$
basis set. As shown in 
Table~\ref{tab:GAFQMC-vs-CCSD(T)},
the AF QMC and CCSD(T) absolute total energies agree to within
$0.07$ eV, and their ionization energies agree to within $0.04$ eV.
These results indicate that the
intrinsic error
due to the phaseless approximation in the above planewave calculations is likely
also small, consistent with previous experience with phaseless AF QMC\cite{sz-hk, AFQMC-CPC2005,
Al-Saidi_TMO, Al-Saidi, Purwanto2004}.
This 
suggests that errors due to the use of pseudopotentials are largely
responsible for the deviations in ionization energies noted above.

\begin{table}[!htbp]
\newcommand\blank[1] {\multicolumn{#1}{c}{}}
\caption{\label{tab:GAFQMC-vs-CCSD(T)}
Calculated total energies ($E$) and ionization energies (IP) for Cl,
using AF QMC with a Gaussian basis together with the corresponding
CCSD(T) results, using identical basis sets.  The \ECPpsp{} is used
with a double-zeta quality $(7s7p2d)$ Gaussian basis set.  All
energies are in eV. Since CCSD(T) is known to be accurate for atoms,
differences compared to Gaussian-based AF QMC provide an estimate of
errors due to the phaseless approximation in AF QMC.
}
\setlength\tabcolsep{3pt}
\begin{tabular}{l r@{.}l r@{.}l  c  r@{.}l r@{.}l}

\hline
\hline

& \multicolumn{4}{c}{AF QMC} &
& \multicolumn{4}{c}{CCSD(T)}
\\
\cline{2-5}
\cline{7-10}
& \multicolumn{2}{c}{$E$} & \multicolumn{2}{c}{IP} & \quad
& \multicolumn{2}{c}{$E$} & \multicolumn{2}{c}{IP}
\\
\hline
\ \ Cl        &\ --403&91(1) & \blank{2}    &&\  --403&96 \ & \blank{2} \\
\ \ Cl$^+$    &\ --391&37(1) & \   12&54(2) &&\  --391&44 \ & \ 12&51 \ \\
\ \ Cl$^{++}$ &\ --368&43(1) & \   35&49(2) &&\  --368&43 \ & \ 35&53 \ \\

\hline
\hline

\end{tabular}
\end{table}

The pseudopotentials used here were generated
using independent-electron HF or DFT mean-field type calculations
of atomic reference systems.
While the transferability of these pseudopotentials to HF or DFT calculations of
molecules or solids is well understood, their accuracy
in many-body calculations is more problematic.\cite{Shirley1993, LeeNeeds}
The dependence of the AF QMC results in
Table~\ref{tab:IP+IIP} on the choice of pseudopotentials is consistent with this.

To estimate the pseudopotential errors in our AF QMC
calculations and to obtain insight into their origin, we have carried out
several additional calculations.
We first performed pseudopotential and all-electron (AE) CCSD(T) calculations to estimate
the transferability of the pseudopotential in a many-body context.
First and second ionization energies
for P, S, and Cl atoms were calculated using both methods.
The coupled-cluster calculations were performed using the \GAUSSIAN[98]
and \GAUSSIAN[03] packages\cite{Gau98,Gau03}.
To eliminate basis set convergence errors, we performed
a series of AE calculations using the aug-cc-p$w$CV$x$Z basis
sets\cite{aug-cc-pVXZ, aug-cc-pCVXZ, PNL-basisform}, where $x=$ D, T, Q for
double, triple, and quadruple zeta basis sets, respectively.
The infinite-basis estimate of the total energy, $E_{\infty}$, is then
obtained using extrapolation\cite{Inf-basis}
\eql{eq:E-cbs}
{
    E_{\infty} \approx E_x - b e^{-cx}
    \,,
}
where $x$ is $2, 3, 4$ for double, triple, and quadruple zeta basis
sets, respectively, and $b$ and $c$ are fitting parameters.
We then take the difference of the extrapolated energies as the ionization
potential shown in Table~\ref{tab:CCSDT-IP}.
For the pseudopotential calculations, the \ECP{} ECP is used.
Here we use the aug-cc-pV$x$Z basis sets\cite{aug-cc-pVXZ, PNL-basisform}
that are fully uncontracted, and again we use the extrapolation scheme
in Eq.~(\ref{eq:E-cbs}).
\COMMENTED{
We verified that the effect of uncontraction on the AE calculations is small.
With and without contraction, AE calculations
resulted in consistent $E_{\infty}$ estimates, with
the change in the absolute energy $\simeq 0.02$ eV,
while the change in ionization energies was a factor of
five to ten smaller.
}

\begin{table}[!htbp]
\caption{\label{tab:CCSDT-IP}
Estimates of pseudopotential errors:
CCSD(T) results for the first and second ionization energies of
P, S, and Cl.
All energies are in eV.
\ECP{} and AE results are shown together with
the error in the ionization energy due to the pseudopotential,
$\Delta\textrm{IP}_\mathrm{psp}$.
}
\setlength\tabcolsep{2.5pt}
\begin{tabular}{c c  r@{.}l r@{.}l r@{.}l r@{.}l  c  r@{.}l r@{.}l r@{.}l r@{.}l}

\hline
\hline
&
& \multicolumn{8}{c}{IP ($X \rightarrow X^+$)} &
& \multicolumn{8}{c}{IIP ($X \rightarrow X^{++}$)}
\\
\cline{3-10}
\cline{12-19}
 &\quad
& \multicolumn{2}{c}{\ECP{}} & \multicolumn{2}{c}{AE} & \multicolumn{2}{c}{$\Delta\textrm{IP}_\mathrm{psp}$}  & \multicolumn{2}{c}{Expt.} &\quad
& \multicolumn{2}{c}{\ECP{}} & \multicolumn{2}{c}{AE} & \multicolumn{2}{c}{$\Delta\textrm{IIP}_\mathrm{psp}$} & \multicolumn{2}{c}{Expt.}
\\
\hline
P       &&  10&48 &  10&47 &\    0&01 &  10&49 &&  30&13 &  30&18 &\  --0&05 &  30&26 \\
S       &&  10&24 &  10&29 &\  --0&05 &  10&36 &&  33&64 &  33&64 &\    0&00 &  33&67 \\
Cl      &&  12&92 &  13&05 &\  --0&13 &  12&97 &&  36&51 &  36&77 &\  --0&26 &  36&78 \\

\hline
\hline

\end{tabular}
\end{table}

The CCSD(T) results for AE and \ECP{} calculations are presented
in Table~\ref{tab:CCSDT-IP}.
AE CCSD(T) ionization energies are seen to be in excellent agreement with
experimental values.
[We note that AE calculations using aug-cc-pCV$x$Z basis set (a variation
of aug-cc-p$w$CV$x$Z) yield almost identical result, where
the estimated IP differs by $\lesssim 0.03$ eV, and IIP by $\lesssim 0.07$ eV.]
The difference between the AE and \ECP{} results provides an estimate of
the error due to the pseudopotential in many-body calculations.
The results in Table~\ref{tab:CCSDT-IP} indicate that
the \ECP{} pseudopotential tends to
underestimate the ionization energies.
The rms error attributable to the pseudopotential is about
$0.12$ eV.
This is not negligible compared to the overall
AF QMC error, which is
$\langle\Delta\textrm{IP}(\textrm{\ECP{}})\rangle_{\mathrm{rms}} = 0.25$
eV for P, S, and Cl first and second ionization energies.

An additional possible source of pseudopotential error in the planewave AF QMC
calculations is the use of the fully nonlocal, separable
Kleinman-Bylander (KB) construction of the pseudopotential.
For example, the \ECP{} ECP is defined in the usual semilocal form, which is used
in quantum chemistry programs:
\eql{eq:PSP-semiloc}
{
    \Veiop^{\textrm{(SL)}}(\mathbf{r})
  = 
    \sum_{l,m} \ket{Y_{lm}} V_l(r) \bra{Y_{lm}}
    \,.
}
where 
$V_l(r)$ is the angular-momentum-dependent potential.
For efficient use in the planewave calculations, it is common
to express this pseudopotential in the fully nonlocal separable KB form
shown in Eq.~(\ref{eq:PSP-KB}).
While the KB and semilocal forms are identical when they act on the reference
atomic state, the KB form can differ for other states.

To investigate the effect of pseudopotential KB formation,
we perform LDA \cite{PerdewWang}
calculations with the \ECPpsp{}:
(1) planewave basis calculations with the KB form of the \ECP{}
    using the \ABINIT{} package 
    (calculations are converged with respect to $\Ecut$ and box size), and
(2) local basis calculations with the semilocal form of the \ECP{} using
    \GAUSSIAN[98]
    (again we use the sequence of aug-cc-pV$x$Z basis sets to extrapolate
    to the infinite basis limit).
The OAL pseudopotential was converted to the KB form using pseudo-orbitals
obtained in an LDA calculation for the neutral atom. For the purpose of constructing
the fully nonlocal projectors, the effects of using LDA rather than HF pseudo-orbitals are
not expected to to be significant.
In both methods, the total-energies are converged to within $0.5$ mHa ($\approx 0.01$ eV).
Table~\ref{tab:KB-test} presents the results for 
the Cl ionization energies.
The \ECPpsp{} expressed in the KB form tends to
underestimate the LDA total energies, with the discrepancy increasing with the
ionization state.
Ionization energies are thus underestimated by up to $0.15$ eV.
The same trend is
observed in the calculated planewave AF QMC ionization energies compared to experiment, 
which indicates that the KB form may contribute errors 
of the order of $0.1-0.2$ eV for Cl using the \ECPpsp{}.
It is clear that the quality of the pseudopotential is 
crucial for obtaining accurate results with AF QMC.

\begin{table}[!htbp]
\newcommand\blank[1] {\multicolumn{#1}{c}{}}
\caption{\label{tab:KB-test}
LDA calculations of chlorine energies, the Kleinman-Bylander (KB) and
semilocal (SL) forms of the OAL pseudopotential.
The programs used are indicated in parantheses, and 
all energies are in eV.
}
\setlength\tabcolsep{2pt}
\begin{tabular}{l r@{.}l r@{.}l  c  r@{.}l r@{.}l  c  r@{.}l}

\hline
\hline

& \multicolumn{4}{c}{Kleinman-Bylander} &
& \multicolumn{4}{c}{Semilocal}
\\

& \multicolumn{4}{c}{(\ABINIT)} &
& \multicolumn{4}{c}{(\GAUSSIAN[98])}
\\
\cline{2-5}
\cline{7-10}
System
& \multicolumn{2}{c}{$E$} & \multicolumn{2}{c}{IP} & \quad
& \multicolumn{2}{c}{$E$} & \multicolumn{2}{c}{IP} & {\ }
& \multicolumn{2}{c}{$(E^{\mathrm{KB}} \! - \! E^{\mathrm{SL}})$}
\\
\hline
\ \ Cl         & \  --403&798  & \blank{2}  && \  --403&752 & \blank{2} && \ \ \  --0&05 \\
\ \ Cl$^+$     & \  --390&842  & \  12&956  && \  --390&701 & \  13&051 && \ \ \  --0&14 \\
\ \ Cl$^{++}$  & \  --367&217  & \  36&581  && \  --367&021 & \  36&731 && \ \ \  --0&20 \\

\hline
\hline

\end{tabular}
\end{table}

These test calculations are limited in scope.
Small systematic errors may well be still present, for example
from the planewave size-extrapolations, 
from Gaussian basis set extrapolations, and from approximations inherent 
in coupled cluster calculations at the CCSD(T) level.
They suggest, however, a
rather consistent picture for understanding the AF QMC results in 
Table~\ref{tab:IP+IIP}. 
Pseudopotential errors due to different origins appear to be the main cause 
for the discrepancies with experimental values. When such errors are 
removed, it seems that the accuracy of the planewave AF QMC is at the 
level of 0.1 eV.

\begin{table}[!hbtp]

\caption{\label{tab:PSP-quality}
Phosphorus ionization and dissociation energies for P$^+$, P$^{++}$, and
P$_2$ computed using LDA, HF, AF QMC, and CCSD(T) methods, shown below in
eV.
AE results are provided to benchmark the pseudopotentials.
The numbers shown in boldface are those closest to the AE
results (for LDA and HF) or the experimental values (for AF QMC).
}

\newcommand\Rd[2] {\textbf{#1}&\textbf{#2}}
\newcommand\blank[1] {\multicolumn{#1}{c}{}}
\renewcommand\arraystretch{1.2}

\setlength\tabcolsep{5pt}
\begin{tabular}{ l r@{.}l r@{.}l r@{.}l r@{.}l }

\hline
\hline
PSP &
\multicolumn{2}{c}{LDA} &
\multicolumn{2}{c}{HF} &
\multicolumn{2}{c}{AF QMC} &
\multicolumn{2}{c}{CCSD(T)} \\
\hline
\multicolumn{5}{l}{IP (P $\rightarrow$ P$^+$)} &
\blank{2} & expt = 10&49
\\
OPIUM   &  \Rd{10}{57}  &      9&97    &     10&74(6)   &\blank{2} \\
\ECP{}  &      10&41    &  \Rd{9}{94}  & \Rd{10}{61(3)} &   10&48  \\
AE      &      10&53    &      9&91    & \blank{2}      &   10&47  \\
\hline
\multicolumn{5}{l}{IIP (P $\rightarrow$ P$^{++}$)} &
\blank{2} & expt = 30&26
\\
OPIUM   &  \Rd{30}{42}  &     29&41    &     30&79(6)   &\blank{2} \\
\ECP{}  &      29&97    & \Rd{29}{11}  & \Rd{30}{28(6)} &   30&13  \\
AE      &      30&37    &     29&08    & \blank{2}      &   30&18  \\
\hline
\multicolumn{5}{l}{$D_e$ (P$_2 \rightarrow 2$P)} &
\blank{2} & expt = 5&08
\\
OPIUM   &   \Rd{5}{97}  &  \Rd{1}{74}  & \Rd{5}{19(16)} &\blank{2} \\
\ECP{}  &       5&29    &      0&98    &     3&88(8)    &    4&39  \\
AE      &       6&18    &      1&65    & \blank{2}      &    4&98\footnote{Frozen-core calculation.}  \\
\hline
\hline
\end{tabular}
\end{table}

Since the quality of a pseudopotential within HF and LDA can be
easily determined (by its ability to reproduce AE results),
it is interesting to see how well this correlates with the
performance of the pseudopotential in a many-body calculation.
In Table~\ref{tab:PSP-quality} we show various energies
obtained from LDA, HF, AF QMC, and CCSD(T) calculations in phosphorus.
(Similar trends also hold for sulfur and chlorine.)
Results are shown using the OPIUM and \ECP{} pseudopotentials, together with
AE results for independent-electron and CCSD(T) methods.
In the LDA calculations, the LDA-based OPIUM potential performs uniformly
better for all quantities.
In the HF calculations, the HF-based \ECP{} pseudopotential performs
well, except for the dissociation energy. 
(As mentioned earlier, the \ECPpsp{} lacks a $d$ projector in its
construction, which results in poor molecular energies.)
Overall, the HF method appears to give a better indication of 
pseudopotential performance with AF QMC.
In the chlorine second ionization energy, for example, HF predicts that
the OPIUM pseudopotential performs better than the OAL pseudopotential:
IIP$_{\textrm{HF, OPIUM}} = 34.37$ eV,
IIP$_{\textrm{HF, \ECP{}}} = 34.30$ eV,
compared to
IIP$_{\textrm{HF, all-electron}} = 34.36$ eV.
AF QMC results in Table~\ref{tab:IP+IIP} show a similar trend.
We also note in Table~\ref{tab:IP+IIP} that
the LDA-based OPIUM pseudopotentials tend to overestimate the ionization
energies, while the OAL pseudopotentials do the opposite.
The tabulated rms averages suggest that the performance of AF QMC with the
LDA-generated OPIUM
pseudopotential varies more widely across different atomic species, especially
in the second ionization energies.
The HF-generated \ECP{} ECP, on the other hand, performs more consistently and yields better
agreement with experiment in the majority of species studied here.
Testing pseudopotentials using
Hartree-Fock calculations may, therefore, be a useful predictor of their
performance in the many-body
AF QMC method.

\section{Summary}
\label{sec:summary}

We have presented electronic structure calculations in atoms and 
molecules using the phaseless AF QMC method with a planewave basis 
and norm-conserving pseudopotentials. 
Various algorithmic issues and characteristics were described and 
discussed in some detail, and 
we have illustrated how the AF QMC method can be implemented by 
utilizing standard DFT planewave techniques. 
The structure of the AF QMC calculation is an independent collection
of random walker streams. Each stream resembles an LDA calculation, which 
makes the overall computational scaling of the method similar to LDA calculation 
with a large prefactor. This makes the AF QMC approach more efficient than explicit many-body
methods.
All of the reported results were obtained using single-determinant
trial wave functions, directly obtained from either LDA or HF calculations.
This reduces the demand for wave
function optimization in QMC and is potentially 
an advantage. The method also offers a different route to the 
sign problem by carrying out the random walks in Slater determinant space. 
Because our method is based in Slater determinant space, any single-particle
basis can be used. 

Results for the dissociation and ionization energies of
second-row atoms and dimers in Group 3A-7A, Al, Si, P, S, Cl,
as well as the As$_2$ dimer, were presented using the planewave-based 
phaseless AF QMC method.
The effects of the phaseless approximation in AF QMC were studied, and 
the accuracy of the pseudopotentials were examined.
Comparisons were made with experiment and with results from other
methods including the DMC and CCSD(T).
Errors due to the phaseless approximation were found to be small, but
non-negligible pseudopotential errors were observed in some cases.
In addition to pseudopotential errors that arise due to their construction in
mean-field type DFT or HF calculations,
possible errors in ionization energies arising from the separable Kleinman-Bylander
form of the pseudopotentials were also observed.
With the appropriate pseudopotentials, the 
method yielded consistently accurate results.

\begin{acknowledgments}

We acknowledge the support of DOE through the
Computational Materials Science Network (CMSN), 
ARO (grant no.~48752PH), NSF (grant no.~DMR-0535529),
and ONR (grant no.~N000140510055).
Computing was done in the Center of Piezoelectric by Design (CPD) and
National Center for Supercomputing Applications (NCSA).
W.~P.\ would also like to thank Richard Martin,
Wissam Al-Saidi, and Hendra Kwee for many
fruitful discussions.

\end{acknowledgments}


\bibliography{bibdb}

\end{document}